9-12-2014

# Dynamic Switching of GOP Configurations in High Efficiency Video Coding (HEVC) using Relational Databases for Multi-objective Optimization

Gangadharan Esakki

Gangadharan Esakki

*Candidate*

Electrical and Computer Engineering

*Department*

This thesis is approved, and it is acceptable in quality and form for publication:

*Approved by the Thesis Committee:*

Dr. Marios Pattichis, Chairperson

Dr. Daniel Llamocca

Dr. Yuebing Jiang

# Dynamic Switching of GOP Configurations in High Efficiency Video Coding (HEVC) using Relational Databases for Multi-objective Optimization

**By**

**Gangadharan Esakki**

**B.E., Electronics & Communication Engineering, Anna University, 2009**

THESIS

Submitted in Partial Fulfillment of the
Requirements for the Degree of

**Master of Science**

**Computer Engineering**

The University of New Mexico
Albuquerque, New Mexico

July 2014

ii

# DEDICATION

This work is dedicated to my beloved mother Subbulakshmi Esakki and my father Esakki Sankarapandian. Without their valuable support and effort I couldn't have achieved more from a humble background. Thank you for always being on my side.

# ACKNOWLEDGMENTS

I would like to thank my advisor and committee chair Dr.Marios Pattichis for his guidance, support and encouragement during my Master's program and also for being an outstanding teacher in the classroom.

I also want to thank Dr.Daniel Llamocca and Dr.Yuebing Jiang for accepting to be a part of the committee in my thesis defense.

I am grateful to Department of Spanish and Portuguese for employing me as computer support when I first came to UNM. Also, I want to thank Dr.Nasir Ghani and Dr.Mark Gilmore for giving me a teaching assistantship when I mostly needed a funded support from the Department of Electrical and Computer Engineering.

Also I would like to thank my university professor Muralikrishnan in India for encouraging and advising me during the early stages of my career. And also I need to mention my special thanks to my friend Venkatesh Jatla and sincere thanks for the great help and support I received from all of my friends, Hindu Temple Society of New Mexico (HTSNM) and Sani Yoga Studio.

# Dynamic Switching of GOP Configurations in High Efficiency Video Coding (HEVC) using Relational Databases for Multi-Objective Optimization

By

Gangadharan Esakki

B.E., Electronics & Communication Engineering, Anna University, 2009
M.S., Computer Engineering, University of New Mexico, 2014

## Abstract

Our current technological era is flooded with smart devices that provide significant computational resources that require optimal video communications solutions. Optimal and dynamic management of video bitrate, quality and energy needs to take into account their inter-dependencies. With emerging network generations providing higher bandwidth rates, there is also a growing need to communicate video with the best quality subject to the availability of resources such as computational power and available bandwidth. Similarly, for accommodating multiple users, there is a need to minimize bitrate requirements while sustaining video quality for reasonable encoding times.

This thesis focuses on providing an efficient mechanism for deriving optimal solutions for High Efficiency Video Coding (HEVC) based on dynamic switching of GOP configurations. The approach provides a basic system for multi-objective optimization approach with constraints on power, video quality and bitrate. This is accomplished by utilizing a recently introduced framework known as Dynamically

Reconfigurable Architectures for Time-varying Image Constraints (DRASTIC) in HEVC/H.265 encoder with six different GOP configurations to support optimization modes for minimum rate, maximum quality and minimum computational time (minimum energy in constant power configuration) mode of operation. Pareto-optimal GOP configurations are used in implementing the DRASTIC modes.

Additionally, this thesis also presents a relational database formulation for supporting multiple devices that are characterized by different screen resolutions and computational resources. This approach is applicable to internet-based video streaming to different devices where the videos have been pre-compressed. Here, the video configuration modes are determined based on the application of database queries applied to relational databases. The database queries are used to retrieve a Pareto-optimal configuration based on real-time user requirements, device, and network constraints.

# TABLE OF CONTENTS

## List of Figures

# List of Tables

# Chapter 1

# Introduction

In our current technical era, image and video communications are omnipresent. With growing network speeds, the demands for image and video communications have increased substantially. An efficient mechanism for adaptively optimizing video content needs to consider real-time constraints on bandwidth, energy and quality of the image/video. To effectively address such issues, we need to reconsider the standard use of Rate-Distortion (RD) optimization in terms of required video quality, available bandwidth, and energy. Furthermore, it is clear that these requirements can vary over time due to both network variations, availability of a power source, or user interest in video content. Beyond video communications, users can also impose their own requirements on video quality, user bandwidth, and encoding times.

The thesis approach is based on the use of Dynamically Reconfigurable Architecture for Time-varying Image Constraints (DRASTIC) that covers software-only configurations [1] to achieve Pareto optimization over a set of general modes that include: (i) maximum image quality, (ii) minimum energy and (iii) minimum bitrate over a set of opposing constraints to guarantee best performance. All these modes are presented in a dynamic encoding framework that allows users (or the network) to impose time varying constraints and optimization requirements on different video segments. For optimal encoding, the thesis investigates the use of different Group of Pictures (GOP) configurations of the emerging High Efficiency Video Coding (HEVC). Optimal

GOP configurations are tested using an HEVC training video that is used to demonstrate dynamic switching for the minimum bitrate mode and the maximum video quality mode. The thesis also introduces a relational-database framework for retrieving Pareto-optimal configurations for different devices and networks.

## 1.1 Motivation

The motivation for the current thesis comes from the need to develop a top-down design approach for adapting to real-time varying constraints for video communications. This top-down approach is based on the use of GOP configurations for HEVC. The database approach is motivated by the need to effectively describe Pareto-optimal configurations for different devices and network configurations.

## 1.2 Thesis Statement

The thesis of this research is that we can use GOP configurations and different Quantization Parameter (QP) to meet dynamically changing constraints on bitrate, quality, and encoding time. Furthermore, Pareto-optimal solutions can be effectively represented using a relational database framework that can support different devices and networks.

## 1.3 Contribution

The main contribution of this thesis is the use of GOP configurations for implementing DRASTIC modes for video communications. A secondary contribution comes from the use of relational databases to describe complex relationships among the Pareto front, device configurations, and networks.

## 1.4 Summary

Chapter 2 provides brief background information on DRASTIC. Chapter 3 provides a summary of HEVC and the use of GOPs. Chapter 4 discusses the implementation of DRASTIC modes by switching of GOP configurations. The relational database formulation and concluding remarks are given in Chapter 5.

# Chapter 2

# Dynamically Reconfigurable Architecture for Time-varying Image Constraints (DRASTIC)

Image and video communications are omnipresent in our current technological era. Right from smart phones to super computers there is a diverse range of consumers with different requirements. With growing network speeds, demands on the image and video processing and communications have exponentially increased resulting in the need for tremendous computational power for any range of systems or devices. Efficient mechanisms for handling these resources and adaptive optimization need to satisfy real-time constraints on bandwidth, energy and quality of the image/video.

To effectively address such issues, we need to reconsider the standard use of Rate-Distortion (RD) optimization and consider the joint optimization of required video quality, available bandwidth, and energy. Furthermore, it is clear that these requirements can vary over time due to both network variations, availability of a power source, or user interest in the video content. Beyond video communications, users can also impose their own requirements on video quality, user bandwidth, and encoding times. For example, YouTube has introduced a new feature in viewing videos online such as changing the resolution of the video being watched. Users now have the option to change the resolution from 240p, 360p, 480p, 720p and Auto mode.

To provide a motivational example, suppose that the user is watching a football match in 720p HD resolution at low speed networks. Due to the lack of bandwidth, the video will likely buffer and will ultimately slow down. In another example, consider the case when a user watches a live streaming of a space shuttle launch in a mobile phone. In this case, video quality requires the visualization of strong motion patterns. In turn, this requires significant computational resources for computing motion vectors on a mobile device. Thus, video streaming will likely consume significant computational resources from a limited energy supply (battery) which is very likely to reduce response times from other running applications.

For effective video communications, an efficacious balance needs to be achieved in order to satisfy the user requirements in real time. This thesis introduces an application [18] of Dynamically Reconfigurable Architectures for Time-Varying Image Constraints (DRASTIC) for video processing systems that can handle real-time constraints through optimal configuration setup [1], [2], [13], [14]. There are four fundamental modes in DRASTIC that totally summarize the requirements for a best performance in the real world. [1]

- **Minimum bit rate mode:** Here the objective is to minimize the bit rate subject to a maximum encoding time and a minimum level of acceptable video quality. For this mode, we note that users can view the video at a higher quality provided that we allow longer processing times to reduce bitrate requirements. Also, the mode can accommodate many users since the bit rate in this mode is minimal.
- **Maximum video quality mode:** In this mode, users prioritize the preservation of the quality of video content is of significant interest to the users. Here the objective is to maximize video quality without exceeding the maximum bandwidth available or maximum encoding time. This mode is suitable for telemedicine applications, or in reviewing special sports events (eg., goals in soccer matches, penalties in sports events, etc).

- **Minimum Energy mode:** This mode takes the encoding time as an estimate of the amount of energy that needs to be expended in the encoding process. Here the objective is to minimize the energy [5] subject to an available bandwidth and a minimum level of acceptable video quality.

- **Typical mode:** This mode optimizes a weighted average of the required encoding time, bitrate, video quality requirements on all of them. This has its own trade-off between those constraints and strives to achieve a balance out of them.

Table 2.1 provides a summary of the modes where a summary of the necessary symbols is given in Table 2.2. In what follows, we will further analyze the different modes.

Table 2.1 DRASTIC modes and Objective functions

| **DRASTIC Modes for Video Communications** | |
|---|---|
| **Mode** | **Constrained Optimization Formulation** |
| Minimum bit rate mode | $\min_{config}$ BPS subject to $(Q > Q_{min})$ & $(T < T_{max})$ |
| Maximum video quality mode | $\max_{config}$ Q subject to $(BPS < BPS_{max})$ & $(T < T_{max})$ |
| Minimum energy mode | $\min_{config}$ T subject to $(Q > Q_{min})$ & $(BPS < BPS_{max})$ |
| Typical mode | max α. Q − β. BPS − γ. T, α + β + γ = 1. subject to $(BPS < BPS_{max})$ & $(T < T_{max})$ & $(Q > Q_{min})$ |

Table 2.2 DRASTIC Objectives and Constraints

| **DRASTIC Objectives and Constraints** | |
|---|---|
| **Objective Function** | **Constraint** |
| T: Encoding Time (sec) | $T_{max}$: Maximum encoding time allowed |
| BPS : Average bits per sample | $BPS_{max}$: Maximum available bits per sample |
| Q : Video quality metric | $Q_{min}$): Minimum acceptable video quality |

This gives an overall picture of the mode of operation in DRASTIC and it is robust in attaining the user requirements for any real-time scenario. The rest of the chapter is organized as follows. In Section 2.1, we describe the framework for DRASTIC. Section 2.1.1 deals with a brief overview of multi-objective optimization approach and its requirements.

## 2.1 Framework for DRASTIC

DRASTIC can be implemented in hardware-only, software-only and a hardware-software joint configuration mode. This thesis focuses on the software-only configurations that extend traditional rate-distortion methods to achieve rate, quality and energy optimization.

The success of DRASTIC relies on the identification of a family of Pareto-optimal configurations [7]. Each Pareto front will have configurations that are developed to get finer control in the rate-quality-energy optimization space.

### 2.1.1 Multi-Objective Optimization Approach

This section gives an overall description of the multi-objective optimization approach. Multi-objective optimization works with different objectives that need to be optimized concurrently. Unlike optimizing over a single objective, in multi-objective optimization [17], different objectives compete or pitted against each other to provide an "optimally best" solution by sacrificing one or more other objectives while optimizing another one [21]. A Pareto surface of solutions summarizes the configurations that are optimal in the multi-objective sense. Constraints reduce the search region to a small portion of the Pareto surface. A scalar objective function is then optimized over the constrained surface.

The three basic objectives to be optimized over the constrained surface are given by:

$$\min_{config} BPS\,(Q, V),\ \max_{config} Q\,(BPS, T),\ \min_{config} T\,(Q, BPS) \qquad (2.1)$$

where *BPS* represents the Bits per sample of the input video *V* taken, *config* represents the encoder configurations assigned during the video encoding, *T* represents the total encoding time which in turn constitutes the energy consumed, *Q* represents the video quality in PSNR dB.

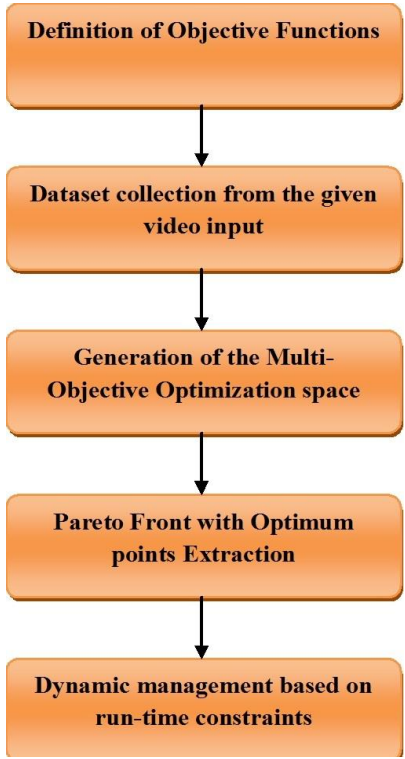
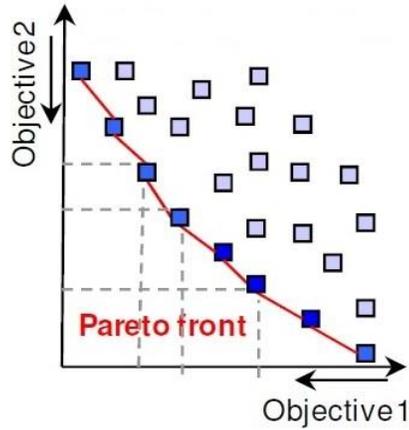

(a)                              (b)

Fig. 2.1(a) Multi-Objective Optimization Framework for DRASTIC. (a) Steps for developing DRASTIC configurations. (b) Example showing a Multi-objective space with a Pareto front. Each point in the graph represents a software configuration mode. The diagram of Fig. 2.1 (b) is taken from [7].

Each one of the objectives in equation 2.1 will result in the selection of an optimal configuration that is selected from the Pareto-front. Here, the Pareto front will be pre-computed using HM12.0 [8]. A flowchart explaining the process is shown in Fig. 2.1(a) and the Pareto front is shown in Fig. 2.1(b) [7]. From the diagram, the non-possible, optimal configuration would be the one that minimizes both objectives. The lower-left point of Fig.2 1(b) represents the impossible configuration that minimizes both objectives. In the most interesting cases, the objectives are inter-dependent (e.g., energy versus performance) that result in trade-offs between them. In this general case, optimality is defined using a multi-objective formulation.

The set of DRASTIC configurations are defined to be Pareto-optimal if an objective cannot be improved upon with sacrificing performance on another objective. In Fig. 2.1(b), the set of Pareto-front points are connected by a red line. This set of Pareto-optimal configurations represents the only set of configurations that are of interest. They represent the best trade-off among the constraints. From Fig. 2.1(b), it is clear that it is not possible to satisfy constraints that require that objectives stay below the Pareto-front. This concept leads to first challenge for developing DRASTIC [13] [17] implementations. Here, note that the Pareto-front determines the objective values that are possible. However, the solutions on the Pareto-front also need to be acceptable. For example, it is not acceptable to have a low-energy configuration that does not deliver sufficient accuracy. Also, to provide flexibility for the optimization, it is necessary to have the configurations provide a sufficiently dense sampling of the Pareto-optimal space so as to allow for efficient implementations for a wide range of constraints.

# Chapter 3

# High Efficiency Video Coding (HEVC)

**High Efficiency Video Coding** (**HEVC**) [4] is a video compression standard, a successor to MPEG-4, H.264/AVC (Advanced Video Coding), that was jointly defined by the ISO/IEC Moving Picture Experts Group (MPEG) and ITU-T Video Coding Experts Group (VCEG) as ISO/IEC 23008-2 *MPEG-H Part 2* and ITU-T *H.265*. MPEG and VCEG established a Joint Collaborative Team on Video Coding (JCT-VC) to develop the HEVC standard. The technical content of HEVC was finalized on January 25, 2013 [22] and the specification was formally ratified as a standard on April 13, 2013. HEVC/H.265 is said to double the data compression ratio compared to MPEG-4 H.264/AVC at the same level of video quality at the same bit rate. Also it can support 4K, 8K UHD and resolutions up to 8192x4320. Still in its nascent stages, several extensions to the technology remain under active development, including range extensions (supporting enhanced video formats), Scalable Video Coding (SVC), and Multi-View Coding (MVC) 3D video extensions.

HEVC was designed to substantially improve coding efficiency compared to H.264/AVC [12], i.e. to reduce bitrate requirements by half with comparable image quality, at the expense of increased computational complexity. HEVC was designed with the goal of allowing video content to have a data compression ratio of up to 1000:1.Depending on the application requirements, HEVC encoders can trade off computational complexity, compression rate, robustness to errors, and encoding delay time. Two of the key features where HEVC was improved compared to H.264/AVC [12] were support for higher resolution video and improved parallel processing [9] methods.

**3.1 Background on HEVC**

The video coding layer of HEVC [4] employs the same "hybrid" approach (inter-/intra-picture prediction and 2D transform coding) used in all video compression standards since MPEG days. Fig. 3.1 depicts the block diagram of a hybrid video encoder, which can create a bit stream conforming to the HEVC standard.

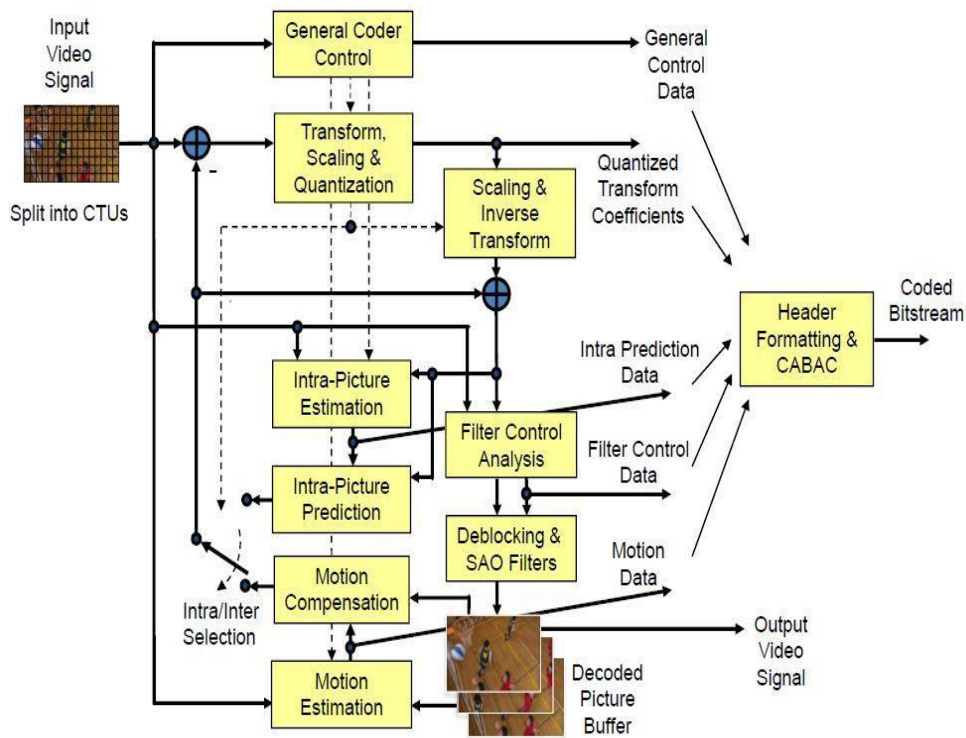

Fig.3.1 Block Diagram of HEVC [4]

An encoding algorithm producing an HEVC compliant bit stream would typically proceed as follows. Each picture is split into block-shaped regions, with the exact block partitioning being conveyed to the decoder. The first picture of a video sequence is coded using only intra-picture prediction. For all remaining pictures of a sequence or between random access points, inter-picture temporally-predictive coding modes are typically used for most blocks. The encoding process for inter-picture prediction consists of choosing motion data comprising the selected reference picture and motion vector (MV) to be applied for predicting the samples of each block. The encoder and decoder generate identical inter prediction signals by applying motion compensation (MC) using the MV and mode decision data, which are transmitted as side information.

The residual signal of the intra or inter prediction, which is the difference between the original block and its prediction, is transformed by a linear spatial transform. The transform coefficients are then scaled, quantized, entropy coded and transmitted together with the prediction information. The encoder duplicates the decoder processing loop such that both will generate identical predictions for subsequent data. Therefore, the quantized transform coefficients are constructed by inverse scaling and are then inverse transformed to duplicate the decoded approximation of the residual signal. The residual is then added to the prediction, and the result of that addition may then be fed into one or two loop filters to smooth out artifacts induced by the block-wise processing and quantization. The final picture representation (which is the duplicate of the output of the decoder) is stored in a decoded picture buffer (DPB) to be used for the prediction of subsequent pictures. In general, the order of the encoding or decoding processing of pictures often differs from the order in which they arrive from the source; necessitating a distinction between the decoding order (bit stream order) and the output order (display order) for a decoder.

### 3.2 Group of Pictures (GOP)

In all video coding standards, GOPs are used to define coding relationships in a sequence of video frames. Generally speaking, all GOP [9], [10] modes start with an I Intra/Key frame that is independently encoded without referencing any other frames. Random access depends on the use of the first key frame within the GOP. Generally, we expect that the use of larger GOP sizes will lead to more compression efficiency. Effective DRASTIC control can be accomplished using a variable GOP structure that supports different inter-mode prediction modes. For the purposes of this thesis, the GOP size will be adaptively changed using the HM12.0 Reference Software [8].

HEVC has two primary ways of arranging GOPs: (a) a closed GOP that uses instantaneous decoding refresh (IDR) picture types that do not have dependencies outside of the GOP, and (b) and Open GOP that uses clean random access (CRA) picture types [refer the HM12.0 technical draft]. HEVC [4] like other video coding standards primarily utilizes three frame types: I (Intra), P (Predictive), and B (Bi-Predictive) types as shown in figure 3.2a [8]. More specifically, we have:

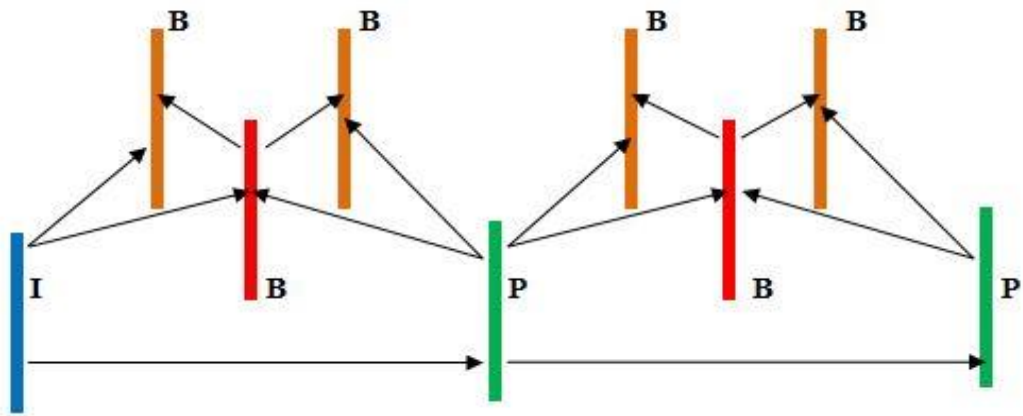

**Fig 3.2a.** An Example of a Closed GOP Structure in HEVC [?refer HM12.0 Reference software draft].

- *I-Pictures:* Intra (I) pictures, also known as *reference* or *key frames* contain all the necessary video information for decoding, without referencing any other frames. Typically, I pictures require more bits to encode than other frame types. Each GOP has an I picture that may or may not be the first frame in the coding video sequence (CVS). I pictures are encoded using angular spatial prediction with modes ranging from 0 to 34 and *take lesser time to encode* compared to P or B pictures.
- *P-Pictures:* Predicted (P) pictures are coded using inter-prediction with one motion-compensation signal per PU (i.e. uni-direction) based on the availability of a closest preceding I or P. P pictures only use reference picture list 0 (L0) and P pictures take *moderate amount of time to encode* compared to I-frames.

- ***B-Pictures***: Bi-directional (B) pictures are coded using inter-prediction with two motion-compensated signals per PU (i.e. bi-direction) based on the availability of I and P pictures that come before and after them. B pictures use reference picture list 0 (L0) and list 1 (L1). B pictures take longer time to encode compared to both I and P-pictures.

GOPs can be specified and defined by three constituents: (i) I, P, and B picture design structure, (ii) the number of video frames used in the GOP, and (c) the type of GOP ("open" or "closed"). The usual GOP designs in a consumer media market such as in DVD, Blue-ray where picture quality is given primary importance are IBP and IBBP. Sometimes, videos are intra-encoded using only I frames to allow for high speed decoding and viewing. The length of the GOP also varies depending upon the application. Longer GOPs [10], [14] are encoded efficiently but do not capture the motion or movement when the camera is under fast transitions, zooming or panning. Whereas the smaller GOPs work very well with videos with higher motion requirements, at the expense of compression efficiency.

### 3.2.1 Instantaneous Decoding Refresh (IDR) Picture Type

As stated in [reference tech. draft], IDR requires that video frames are partitioned into I slices. The first I slice in IDR may be the first in the original video sequence or may come at a later time in the sequence. An IDR picture type example is shown in fig 3.2a where the frames decoded refer to the IDR picture, and cannot refer to frames preceding the IDR. Simply put, it forms a closed loop structure making it more self-reliant and independent. Additionally, the GOP uses cyclic or dyadic arrangement of B or P pictures forming a hierarchy of reference pictures, increasing the encoding time and complexity while the GOP structure allows for more robust error control.

### 3.2.2 Clean Random Access (CRA) Picture Type

Similar to IDR, a CRA picture contains only I slices that can come from different times in the video sequence. On the other hand, a CRA picture belongs to open GOP structure as shown in fig. 3.2b The Figure shows that the frames in the GOP can refer to other pictures preceding the CRA picture in the decoding order. This kind of arrangement makes the compression more efficient by requiring lower bitrates while facilitating random access. For example, when watching a movie in YouTube, if the user wants to fast forward and seek a specific timeline, the video can be encoded using an open GOP that provides the random access feature that allows decoding to start at the nearest CRA picture that is then used for decoding the pictures that refer to it.

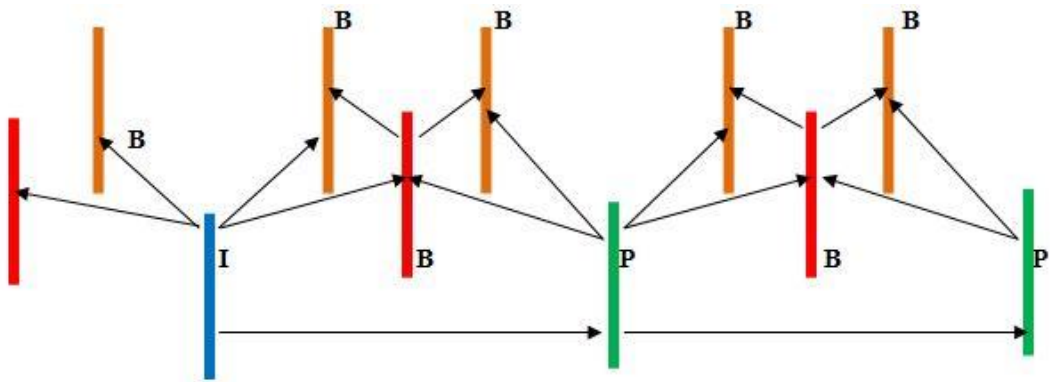

**Fig 3.2b** An Example of an Open GOP Structure.

### 3.3 Deblocking and Sample Adaptive Offset Filtering

In HEVC [refer HEVC main paper], two filters are used: i) Deblocking Filtering (DBF), and ii) Sample Adaptive Offset (SAO) filtering. Both filters are used in the frame reconstruction loop to remove the artifacts induced because of block partitioning in the video coding. Primarily, this thesis is built upon proper utilization of these filters as they play a key role in GOP switching and affect the quality of the reconstructed video frames. The Deblocking filter in HEVC is applied to the block boundaries of each picture where often the artifacts appear at lower bitrates (similar to H.264 [12], [15]). Beyond deblocking filtering used in H.264, HEVC introduced the use of SAO. Based on our experimental setup, the HEVC HM12.0 [8] encodes videos by adaptively applying these deblocking filters by turning them ON and OFF.

Deblocking in HEVC can be applied to all samples that are in the boundaries of TU and PU for blocks of sizes 8x8 or higher. Instead, H.264/AVC uses deblocking every 4x4 grid edge. HEVC does not apply deblocking on picture, tile, and slice boundaries. Here, we note that tiles were newly introduced to HEVC. A Tile decomposition involves the use of independent, rectangular regions that can provide better support for parallelism as opposed to the standard use of slices. The deblocking algorithm works in the following manner. First, the algorithm determines the filter strength in an 8x8 grid for both the vertical and horizontal edges and then computes thresholds that depend on the boundary filter strength and quantization parameter (QP). The deblocking filter in HEVC has the following boundary strengths: 2 (strong), 1 (weak) and 0 (no deblocking), similar to H.264/AVC that supports five boundary strengths. One of the advantages of the deblocking filter in HEVC is that it works both the edges independently, which enables a parallelized implementation. In theory, it would be possible to perform vertical edge filtering with one thread per 8-pixel column in the picture.

### 3.4 Sample Adaptive Offset (SAO)

The additional module introduced in the loop filtering is Sample Adaptive Offset or SAO [4]. During reconstruction of the samples, the quantization error can lead to significant edge artifacts. SAO is an optional filtering tool to be applied after deblocking in order to reduce these errors. This filter can be optionally turned off for both the luma and chroma samples or can be applied to each of them separately. Unlike regular deblocking, SAO is adaptively applied to all pixels. There are two types of SAO: 1. Edge type where the sample offset depends on the edge mode and 2. Band type where the offset depends on the shape amplitudes in terms of pixel bands ranging from 0 to 255.

### 3.5 Encoder Configuration Modes

The HM 12.0 reference software encoder [9] has three primary configurations. These configurations [23], [24] include: all intra (AI) mode, random access (RA) mode, and a low delay (LD) mode.

**All Intra Mode (AI)**: In this mode, all the frames are encoded only using I slices. It can be an IDR or CRA picture according to the configuration file. And there is no temporal prediction done in this mode as all the pictures are intra-encoded using spatial angular prediction. The quantization parameter for each I picture can be modified within the sequence. Figure 3.5.1 shows an all intra mode representation.

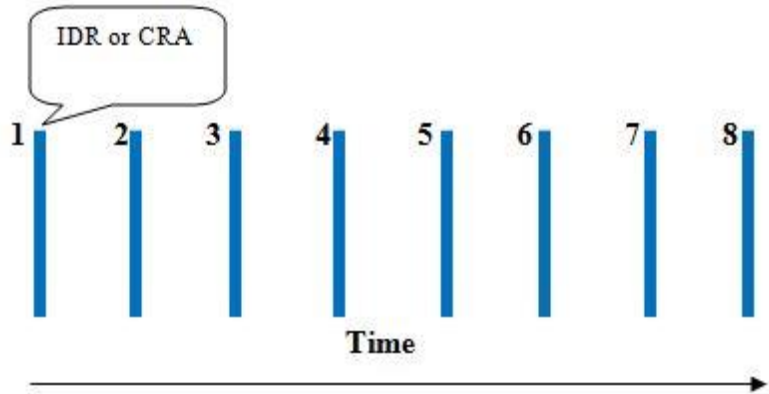

Fig. 3.5.1 All Intra Mode Representation

A typical AI configuration mode file is shown in fig 3.5.2.

```
#======== File I/O =====================
BitstreamFile                : str.bin
ReconFile                    : rec.yuv

#======== Profile ================
Profile                      : main

#======== Unit definition ================
MaxCUWidth                   : 64           # Maximum coding unit width in pixel
MaxCUHeight                  : 64           # Maximum coding unit height in pixel
MaxPartitionDepth            : 4            # Maximum coding unit depth
QuadtreeTULog2MaxSize        : 5            # Log2 of maximum transform size for
                                            # quadtree-based TU coding (2...6)
QuadtreeTULog2MinSize        : 2            # Log2 of minimum transform size for
                                            # quadtree-based TU coding (2...6)
QuadtreeTUMaxDepthInter      : 3
QuadtreeTUMaxDepthIntra      : 3

#======== Coding Structure ==============
IntraPeriod                  : 1            # Period of I-Frame ( -1 = only first)
DecodingRefreshType          : 0            # Random Accesss 0:none, 1:CDR, 2:IDR
GOPSize                      : 1            # GOP Size (number of B slice = GOPSize-1)
#        Type POC QPoffset QPfactor tcOffsetDiv2 betaOffsetDiv2  temporal_id #ref_pics_active #ref_pics reference pictures

#=========== Motion Search =============
FastSearch                   : 1            # 0:Full search  1:TZ search
SearchRange                  : 64           # (0: Search range is a Full frame)
HadamardME                   : 1            # Use of hadamard measure for fractional ME
FEN                          : 1            # Fast encoder decision
FDM                          : 1            # Fast Decision for Merge RD Cost

#======== Quantization ==============
QP                           : 32           # Quantization parameter(0-51)
MaxDeltaQP                   : 0            # CU-based multi-QP optimization
MaxCuDQPDepth                : 0            # Max depth of a minimum CuDQP for sub-LCU-level delta QP
DeltaQpRD                    : 0            # Slice-based multi-QP optimization
RDOQ                         : 1            # RDOQ
RDOQTS                       : 1            # RDOQ for transform skip

#=========== Deblock Filter ============
DeblockingFilterControlPresent: 0           # Dbl control params present (0=not present, 1=present)
LoopFilterOffsetInPPS        : 0            # Dbl params: 0=varying params in SliceHeader, param = base_param + GOP_offset_param; 1=constant params in PPS, param =
base_param
LoopFilterDisable            : 0            # Disable deblocking filter (0=Filter, 1=No Filter)
LoopFilterBetaOffset_div2    : 0            # base_param: -6 ~ 6
LoopFilterTcOffset_div2      : 0            # base_param: -6 ~ 6
DeblockingFilterMetric       : 0            # blockiness metric (automatically configures deblocking parameters in bitstream)

#=========== Misc. =============
InternalBitDepth             : 8            # codec operating bit-depth

#=========== Coding Tools ================
```

Fig. 3.5.2 All Intra Configuration File

**Random Access (RA)**: This mode uses a pyramidal or a dyadic relationship among the B slices. Hierarchical B pictures are used for the coding with a random access picture used every 1s. This mode can be used in digital video broadcasting applications. Figure 3.5.3 shows a random access configuration. Each picture is depicted with a number showing the encoding order. The first I picture of a video sequence is always encoded as an IDR picture and the other intra pictures are encoded as non-IDR intra pictures ("Open GOP"). Since this configuration follows a pyramid like structure, there are several temporal layers within pictures and each of the intermediate pictures is encoded as a B picture. The Intra period, GOP size and type can always be changed and with this Generalized P and B (GPB) structure that can also be used to define the lowest temporal layer that refers to I or GPB picture for inter prediction.

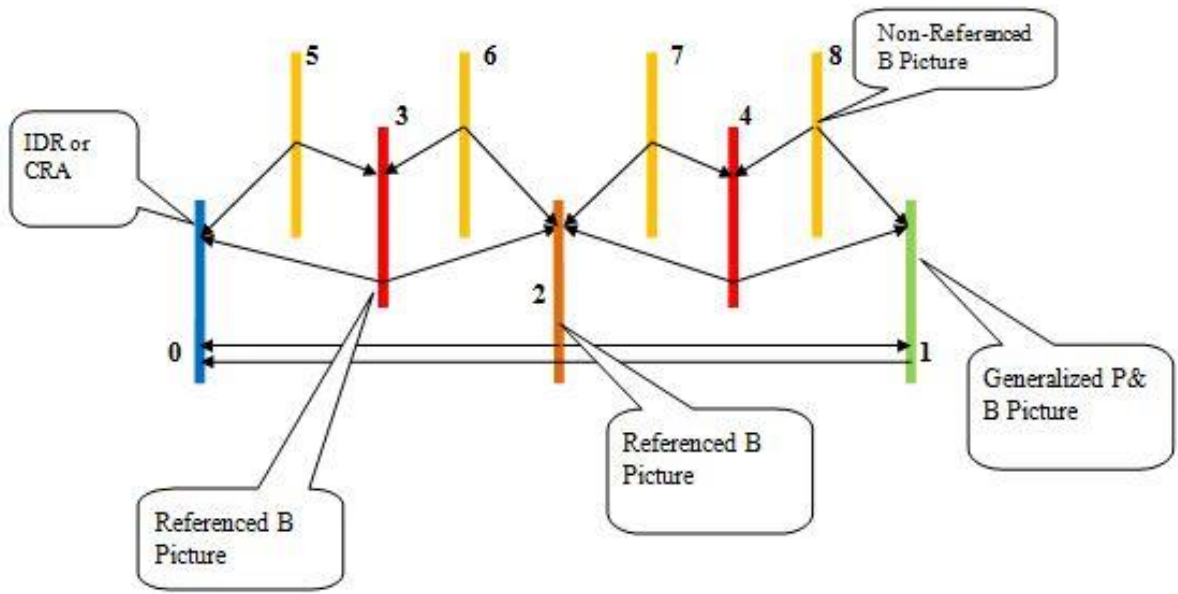

Fig. 3.5.3 Random Access Representation

The other two layers viz., second and third temporal layers have referenced B pictures, and the last temporal layer from the IDR picture contains only the non-referenced B pictures. QP for each inter coded picture is derived by adding a QP offset to QP of Intra coded picture depending on temporal layer. The GOP structure for a RA configuration mode as shown in fig.3.5.4 has a GOP size = 8 and is always associated with the Intra period = 32, that is the occurrence of the IDR picture after consecutive B pictures in the coded video sequence (CVS). It is always ensured that the GOP size is a multiple of the Intra period which includes an IDR or CRA picture if needed for customization.

```
#======== File I/O =====================
BitstreamFile                 : str.bin
ReconFile                     : rec.yuv

#======== Profile ================
Profile                       : main

#======== Unit definition ================
MaxCUWidth                    : 64          # Maximum coding unit width in pixel
MaxCUHeight                   : 64          # Maximum coding unit height in pixel
MaxPartitionDepth             : 4           # Maximum coding unit depth
QuadtreeTULog2MaxSize         : 5           # Log2 of maximum transform size for
                                            # quadtree-based TU coding (2...6)
QuadtreeTULog2MinSize         : 2           # Log2 of minimum transform size for
                                            # quadtree-based TU coding (2...6)
QuadtreeTUMaxDepthInter       : 3
QuadtreeTUMaxDepthIntra       : 3

#======== Coding Structure =============
IntraPeriod                   : 32          # Period of I-Frame ( -1 = only first)
DecodingRefreshType           : 1           # Random Accesss 0:none, 1:CDR, 2:IDR
GOPSize                       : 8           # GOP Size (number of B slice = GOPSize-1)
#        Type POC QPoffset QPfactor tcOffsetDiv2 betaOffsetDiv2 temporal_id #ref_pics_active #ref_pics reference pictures     predict deltaRPS #ref_idcs reference idcs
Frame1:  B    8   1        0.442    0            0              0           4                4         -8 -10 -12 -16         0
Frame2:  B    4   2        0.3536   0            0              0           2                3         -4 -6  4               1       4        5         1 1 0 0 1
Frame3:  B    2   3        0.3536   0            0              0           2                4         -2 -4  2 6             1       2        4         1 1 1 1
Frame4:  B    1   4        0.68     0            0              1           2                4         -1  1  3 7             1       1        5         1 0 1 1 1
Frame5:  B    3   4        0.68     0            0              1           2                4         -1 -3  1 5             1      -2        5         1 1 1 1 0
Frame6:  B    6   3        0.3536   0            0              0           2                4         -2 -4 -6 2             1      -3        5         1 1 1 1 0
Frame7:  B    5   4        0.68     0            0              1           2                4         -1 -5  1 3             1       1        5         1 0 1 1 1
Frame8:  B    7   4        0.68     0            0              1           2                4         -1 -3 -7 1             1      -2        5         1 1 1 1 0

#=========== Motion Search =============
FastSearch                    : 1           # 0:Full search  1:TZ search
SearchRange                   : 64          # (0: Search range is a Full frame)
BipredSearchRange             : 4           # Search range for bi-prediction refinement
HadamardME                    : 1           # Use of hadamard measure for fractional ME
FEN                           : 1           # Fast encoder decision
FDM                           : 1           # Fast Decision for Merge RD cost

#======== Quantization =============
QP                            : 32          # Quantization parameter(0-51)
MaxDeltaQP                    : 0           # CU-based multi-QP optimization
MaxCuDQPDepth                 : 0           # Max depth of a minimum CuDQP for sub-LCU-level delta QP
DeltaQpRD                     : 0           # Slice-based multi-QP optimization
RDOQ                          : 1           # RDOQ
RDOQTS                        : 1           # RDOQ for transform skip

#=========== Deblock Filter ============
DeblockingFilterControlPresent: 0           # Dbl control params present (0=not present, 1=present)
LoopFilterOffsetInPPS         : 0           # Dbl params: 0=varying params in SliceHeader, param = base_param + GOP_offset_param; 1=constant params in PPS, param =
                                            # base_param)
```

Fig. 3.5.4 Random Access Configuration File

**Low Delay (LD):** This configuration slightly differs from the random access mode as there is no picture reordering and only the first frame is encoded using I slices. It is suitable for live streaming and video conferencing applications. Similar to random access, in low-delay coding conditions, the first picture in the CVS is always encoded as an IDR picture. Other pictures in the sequence are coded as encoded as Generalized P and B-pictures (GPB). Since there is no picture reordering as in random access, the GPB uses only the reference pictures, where the picture order or the display order is smaller than the current picture being encoded. In low delay, there are two reference pictures List0 and List1 in the decoded buffers to be used while in the reconstruction of the original picture. The contents of List0 and List1 are identical, and are updated with a sliding-window management process.

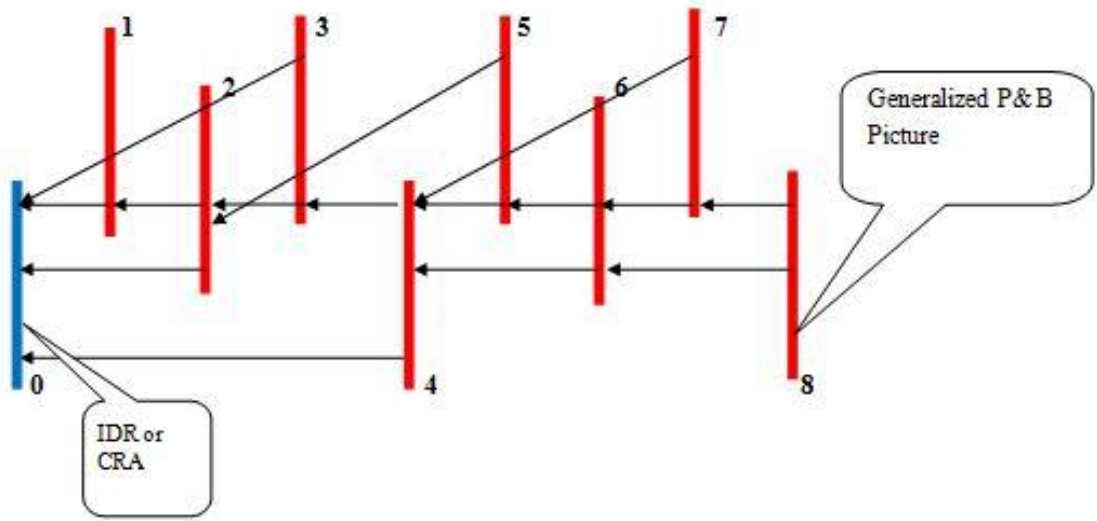

Fig. 3.5.5 Low Delay Representation

The reference pictures shown in the figure 3.5.5 demonstrate that there is no picture reordering or in other words no cyclic (dyadic) relationship among reference B pictures. Each picture is associated with a number representing the encoding order and QP for each picture can be modified corresponding to the QP offset derived from the IDR or CRA picture in the beginning of the GOP. In [HM reference software] there is a provision to use low delay configuration with P-picture encoding using the only the reference picture List0 for inter prediction. Figure 3.5.6 shows a typical low delay configuration with B pictures as the reference frames for the sequence with GOP size =4 and the Intra Period = -1, meaning that the first frame is encoded as an I picture. Also, recall that the GOP size is also customizable but should always be a power of 2 and a function of the reference pictures. The use of a smaller number of reference pictures is used to decrease the temporal redundancy.

```
#======== File I/O =====================
BitstreamFile                 : str.bin
ReconFile                     : rec.yuv

#======== Profile ================
Profile                       : main

#======== Unit definition ================
MaxCUWidth                    : 64          # Maximum coding unit width in pixel
MaxCUHeight                   : 64          # Maximum coding unit height in pixel
MaxPartitionDepth             : 4           # Maximum coding unit depth
QuadtreeTULog2MaxSize         : 5           # Log2 of maximum transform size for
                                            # quadtree-based TU coding (2...6)
QuadtreeTULog2MinSize         : 2           # Log2 of minimum transform size for
                                            # quadtree-based TU coding (2...6)
QuadtreeTUMaxDepthInter       : 3
QuadtreeTUMaxDepthIntra       : 3

#======== Coding Structure =============
IntraPeriod                   : -1          # Period of I-Frame ( -1 = only first)
DecodingRefreshType           : 0           # Random Accesss 0:none, 1:CDR, 2:IDR
GOPSize                       : 4           # GOP Size (number of B slice = GOPSize-1)
#        Type POC QPoffset QPfactor tcOffsetDiv2 betaOffsetDiv2 temporal_id #ref_pics_active #ref_pics reference pictures predict deltaRPS #ref_idcs reference idcs
Frame1:  B    1   3        0.4624   0            0              0           4                4         -1 -5 -9 -13        0
Frame2:  B    2   2        0.4624   0            0              0           4                4         -1 -2 -6 -10        1        -1        5        1 1 1 0 1
Frame3:  B    3   3        0.4624   0            0              0           4                4         -1 -3 -7 -11        1        -1        5        0 1 1 1 1

Frame4:  B    4   1        0.578    0            0              0           4                4         -1 -4 -8 -12        1        -1        5        0 1 1 1 1

#=========== Motion Search =============
FastSearch                    : 1           # 0:Full search  1:TZ search
SearchRange                   : 64          # (0: Search range is a Full frame)
BipredSearchRange             : 4           # Search range for bi-prediction refinement
HadamardME                    : 1           # Use of hadamard measure for fractional ME
FEN                           : 1           # Fast encoder decision
FDM                           : 1           # Fast Decision for Merge RD cost

#======== Quantization =============
QP                            : 32          # Quantization parameter(0-51)
MaxDeltaQP                    : 0           # CU-based multi-QP optimization
MaxCuDQPDepth                 : 0           # Max depth of a minimum CuDQP for sub-LCU-level delta QP
DeltaQpRD                     : 0           # Slice-based multi-QP optimization
RDOQ                          : 1           # RDOQ
RDOQTS                        : 1           # RDOQ for transform skip

#=========== Deblock Filter ============
DeblockingFilterControlPresent: 0           # Dbl control params present (0=not present, 1=present)
LoopFilterOffsetInPPS         : 0           # Dbl params: 0=varying params in SliceHeader, param = base_param + GOP_offset_param; 1=constant params in PPS, param =
base_param)
LoopFilterDisable             : 0           # Disable deblocking filter (0=Filter, 1=No Filter)
LoopFilterBetaOffset_div2     : 0           # base_param: -6 ~ 6
LoopFilterTcOffset_div2       : 0           # base_param: -6 ~ 6
```

Fig. 3.5.6 Low Delay Configuration File

This chapter provided a summary of all the encoder configuration modes with the GOP structure and their corresponding configuration files. In the next chapter, we will discuss switching between GOPs to support DRASTIC optimization modes.

# Chapter 4

## DRASTIC Based on Dynamic Switching of GOP Configurations

### 4.1 Introduction

This chapter implements DRASTIC modes based on dynamic switching of the GOPs [11], [19]. The chapter focuses on the implementation of maximum video quality and minimum bit rate modes defined within DRASTIC. As mentioned earlier, the approach is based on multi-objective optimization.

To implement DRASTIC based on GOP configurations, we test the standard configurations and introduce two new configurations. The new configurations are: (1) Low delay with GOP size=6 (code=LD6), and (2) Random Access with GOP size=4 or (code=RA4). In total, there are six different GOP configurations and each of them has its own parameters. The most significant parameters for each configuration include: (i) the Quantization parameter (QP), (ii) Decoding refresh type (IDR or CRA), and (iii) the In-loop filters (Deblocking & SAO).

For example, a video of resolution 416x240 (mostly used in today's smart phones and mobile devices), is run with a total 216 of these configurations and produces the output in terms of PSNR, bit rate and encoding time. A Pareto optimization approach is used to select the best configuration from the Pareto front in real-time. The motivation for this research is to follow a top down approach to the multi-objective optimization problem.

## 4.2 Pareto Optimization with GOP modes

For the Pareto optimization we have utilized the development of dynamically reconfigurable video encoding system that can find optimal GOP configurations that can solve the following DRASTIC optimization modes:

*Minimum bitrate mode:* We consider this mode for video communications with limited bandwidth. The optimization mode requires that we select the optimal GOP configuration that solves:

$$\min_{config} BPS \text{ subj. to: } (Q > Q_{min}) \text{ and } (T < T_{max}) \tag{4.1}$$

where: $config$ denotes the HEVC GOP configuration parameters, $BPS$ refers to the number of bits per sample, $Q$ refers to the video quality, $T$ is the encoding time, $Q_{min}$ refers to the minimum acceptable video quality, and $T_{max}$ refers to the maximum encoding time. Here, we note that the encoding time is used as an estimate of the amount of energy that needs to be expended in the encoding process.

*Maximum video quality mode:* We consider this mode when video content is of significant interest to the users. In this case, we want to select the optimal configuration mode that solves:

$$\max_{config} Q \text{ subj. to: } (BPS < BPS_{max}) \text{ \& } (T < T_{max}) \tag{4.2}$$

where: $BPS_{max}$ refers to the maximum available bandwidth and the rest of the symbols are defined as for the minimum bitrate mode.

The basic framework for solving the constraint optimizations problems given by (4.1) and (4.2) require the use of a multi-objective optimization framework as discussed in [4, 1, 2]. Here, the idea is to pre-compute the Pareto front of optimal configurations and then solve (4.1)-(4.2) by selecting the optimal solution along the front.

## 4.3 Methodology

We summarize the basic approach in Fig. 2.1. First, we need to compute the Pareto front over a set of training videos. For each configuration, we compute the average PSNR, bitrate, and encoding time. For computing the Pareto front, we simply reject all configurations for which we can find at-least one other configuration that provides better PSNR while requiring less bitrate and less encoding time. A pseudo code in figure 4.3.1 showing the dynamic GOP [25] encoding time for the DRASTIC mode reconfigurations.

```
InitParetoFront (modes, training-videos)
    apply all modes to all of the training-videos
        and estimate (PSNR, bitrate, encoding-time)
        for each mode.
    reject modes that are not Pareto optimal
    return (ParetoFront)

DynamicGOPEncoding(video, modes, ParetoFront)
    frameNum = 1; i=1;
    while frameNum < size-of (video)
        select configuration from ParetoFront that
            satisfies modes(i) requirements
            encode GOP using configuration
            frameNum = frameNum + size-of (GOP)
    end
    return (encodedVideo)
```

Fig.4.3.1 Pseudo code for dynamic GOP encoding subject to DRASTIC mode reconfigurations.

Dynamic GOP encoding requires the use of the Pareto front [17], the desired modes for the different portions of the video, and the actual video to be encoded. For each mode, we select the optimal solution by directly solving (4.1) or (4.2) over the space of the Pareto-optimal configurations.

## 4.4 Experiment Setup

In order to generate the Pareto front, first we need to get the encoder configurations for all modes with the standard and the new GOPs created and additional parameters like QP (Quantization Parameter), DBL (Deblocking filter), SAO (Sample Adaptive Offset), Decoding refresh type (IDR or CRA) etc.. We test the approach on a Basketball Pass video [4], [9] of resolution (416x240) that most smart phones and mobile devices use today with a frame rate (30fps).

### 4.4.1 Encoding Experiment I

In this encoding experiment, we have developed configuration files for the HM encoder [8] with QP values ranging from (22, 27, 31, 32, 33, and 37) and turning ON/OFF the filters DBL and SAO corresponding to the decoding refresh type. This is summarized as a Table I given below. Refer Table.4.1 for configurations.

Table 4.1.GOP parameter encoding first 100 frames for Basketball pass (416x240) video. We have 120 GOP Configurations.

| *Mode* | *QP* | *SAO* | *DBF* | *Decoding Refresh* | *Number of Configs* |
|---|---|---|---|---|---|
| All I | 22, 27, 31, 32, 33, 37 | ON/OFF | ON/OFF | - | 24 |
| RA 8 | 22,27, 31,32, 33,37 | ON/OFF | ON/OFF | IDR/CDR | 48 |
| LD 4 | 22,27, 31,32, 33,37 | ON/OFF | ON/OFF | IDR/CDR | 48 |

In AI mode, the configuration files are defined in terms of a combination of six QP values and filter options to generate a total of 24 configurations. Similarly, for random access (RA8) mode, the parameters vary as a function of decoding refresh type to generate a total of 48 configurations. The same combinations apply to the low delay mode to generate the same number of configurations (48). As shown in the pseudo code, the training phase uses the configurations to generate the Pareto front. Fig 4.4.1a shows the multi-objective optimization space for the first 100 frames of Basketball Pass video of resolution 416x240.

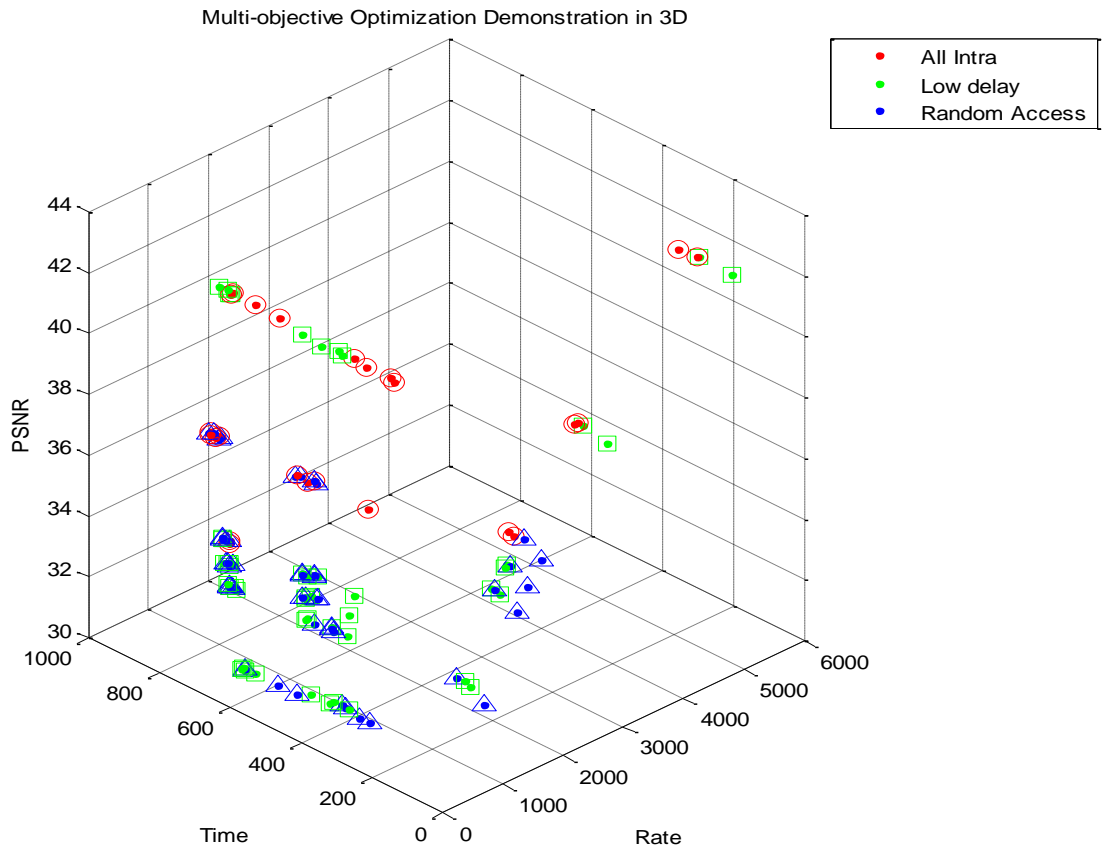

**Fig. 4.4.1a.** Multi-objective optimization space for the first 100 frames of the Basketball Pass video (see text).

From the graph, the red points correspond to All-Intra configurations, green point correspond to the low delay configurations and finally the blue points correspond to random access configurations.

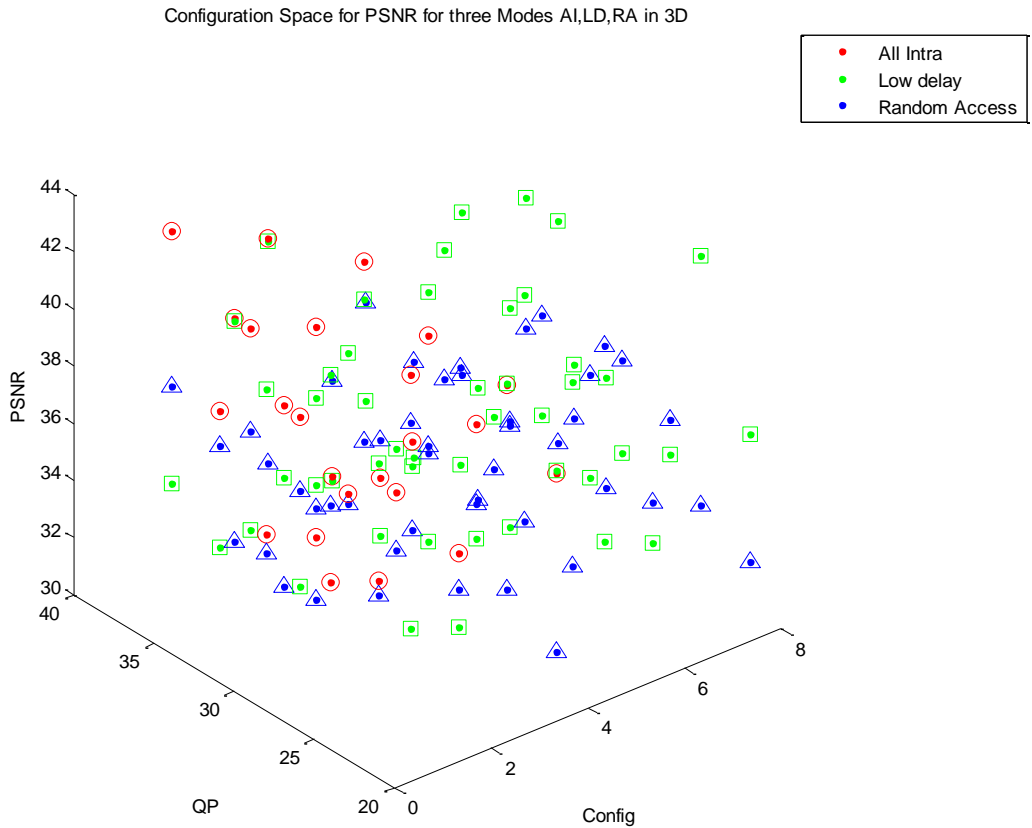

**Fig. 4.4.1b.** PSNR as a function of QP and the GOP configuration.

In Fig. 4.4.1c, we have the encoding time as a function of QP and the GOP configuration number. As we shall see later when examining the Pareto front, AI modes become Pareto-optimal when requiring lower encoding times. From Fig. 4.4.1c, we can deduce that for QP = 22 and Config = 1 the Encoding time = 107 sec which is the fastest of all the modes comes from the AI configuration. But the only disadvantage of AI mode is they are quite lavish with bit rates as they only encode I pictures. Table 4.2 summarizes the results of the AI modes for QP =22.

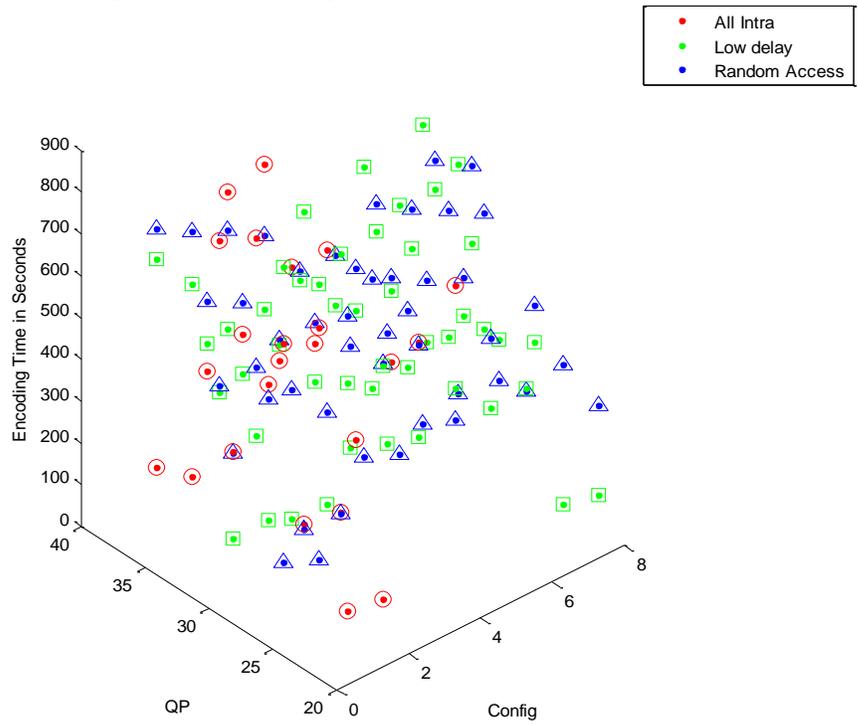

**Fig. 4.4.1c.** Encoding time as a function of QP and the GOP configuration.

**Table 4.2** Summary of AI mode performance for QP=22.

| DBL | SAO | PSNR | TIME | BITRATE |
|---|---|---|---|---|
| ON | ON | 43.0909 | 101.921 | 4866.88 |
| ON | OFF | 43.0451 | 156.127 | 4846.14 |
| OFF | ON | 42.952 | 104.054 | 5111.496 |
| OFF | OFF | 43.0909 | 107.032 | 4866.88 |

Compared to the AI encoder configuration mode, low delay and random access configurations provide better bit rates but at the cost of encoding time and little compromise in PSNR as they involve temporally predicted P & B pictures. Fig. 4.4.1d shows the bitrate as a function of the configuration mode and QP.

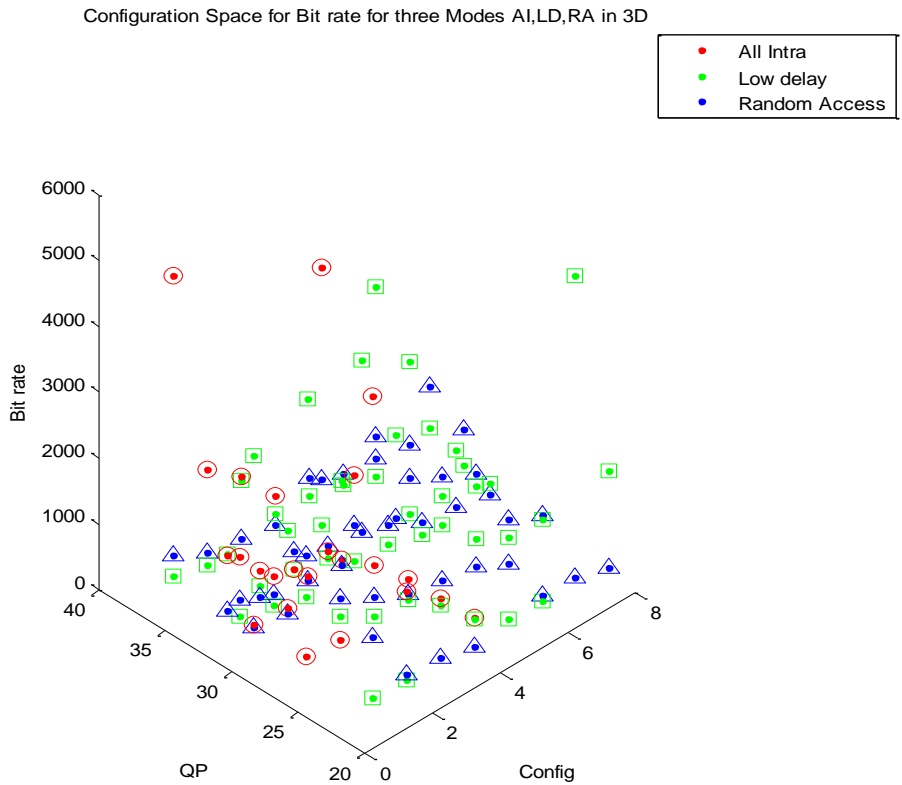

**Fig. 4.4.1d.** Bitrate as a function of QP and the GOP configuration.

From the plot, it is inferred that, AI mode has very higher bit rates than LD or RA modes. The bitrates for random access mode are little lower than low delay mode since in low delay there is no usage of picture reordering and efficient handling of B pictures whereas in random access there are more B pictures for reference frames. Additionally the encoding time for low delay is higher compared to random access mode. In terms of video quality both these modes are more or less the same but it entirely depends upon the application wherein they are used. For QP =22 and decoding refresh type (CRA & IDR) the following table 4.3 summarizes the results for both LD and RA modes.

**Table 4.3** Performance of different configuration modes for QP=22.

LD4, QP = 22, CRA

| DBL | SAO | PSNR | Time | Bitrate |
|---|---|---|---|---|
| ON | ON | 41.1358 | 965.423 | 1127.692 |
| ON | OFF | 41.1358 | 781.953 | 1127.692 |
| OFF | ON | 41.0569 | 959.735 | 1127.256 |
| OFF | OFF | 41.0569 | 675.844 | 1127.256 |

RA8, QP =22, CRA

| DBL | SAO | PSNR | Time | Bitrate |
|---|---|---|---|---|
| ON | ON | 41.407 | 477.995 | 1089.156 |
| ON | OFF | 41.407 | 436.604 | 1089.156 |
| OFF | ON | 41.3507 | 465.97 | 1085.06 |
| OFF | OFF | 41.3507 | 332.199 | 1085.06 |

LD4, QP=22, IDR

| DBL | SAO | PSNR | Time | Bitrate |
|---|---|---|---|---|
| ON | ON | 41.1358 | 1101.94 | 1127.692 |
| ON | OFF | 41.1358 | 1094.985 | 1127.692 |
| OFF | ON | 41.0569 | 1101.344 | 1127.256 |
| OFF | OFF | 41.0569 | 1096.454 | 1127.256 |

RA8, QP=22, IDR

| DBL | SAO | PSNR | Time | Bitrate |
|---|---|---|---|---|
| ON | ON | 41.2959 | 586.055 | 1114.18 |
| ON | OFF | 41.2403 | 532.51 | 1110.384 |
| OFF | ON | 41.2959 | 402.753 | 1114.18 |
| OFF | OFF | 41.2403 | 326.694 | 1110.384 |

In low delay mode, both the CRA & IDR have a very high PSNR = 41.1358 db but the encoding time varies for each of them. Same goes for the random access mode with better PSNR and encoding time. Comparing both the modes, we have some of the values repetitive in PSNR and bitrate but the encoding time makes a difference in the HM encoder software.

## 4.4.1.1 Pareto front for the first 100 frames using standard GOP configurations

The Pareto front is shown in Fig. 4.4.1.1. Out of 120 configurations, we have 50 that are optimal and make up the Pareto front.

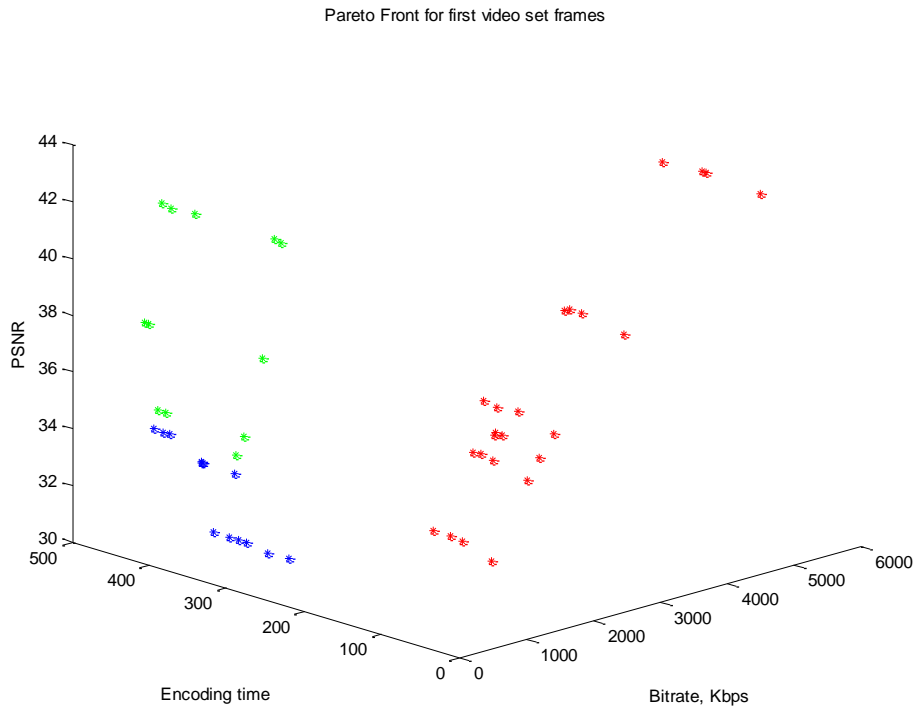

**Fig. 4.4.1.1.** Pareto front for frames 0-100 of Basketball Pass 416x240 video.

The 50 optimal configurations come from all three encoder modes (red - all intra, blue - low delay and green - random access). It is observed from the Pareto front we get a maximum video quality with 43.05 dB which is an All Intra configuration. The minimum bitrate of 142 Kbps is achieved by a random access configuration mode.

## 4.4.2 Encoding Experiment II using both standard and New GOP configurations

In the second encoding experiment, we have developed configuration files for the HM encoder with QP values ranging from (22, 27, 31, 32, 33, and 37) and turning ON/OFF the filters DBL and SAO corresponding to the decoding refresh type. The list of all of the GOP configurations is given in Table 4.4.

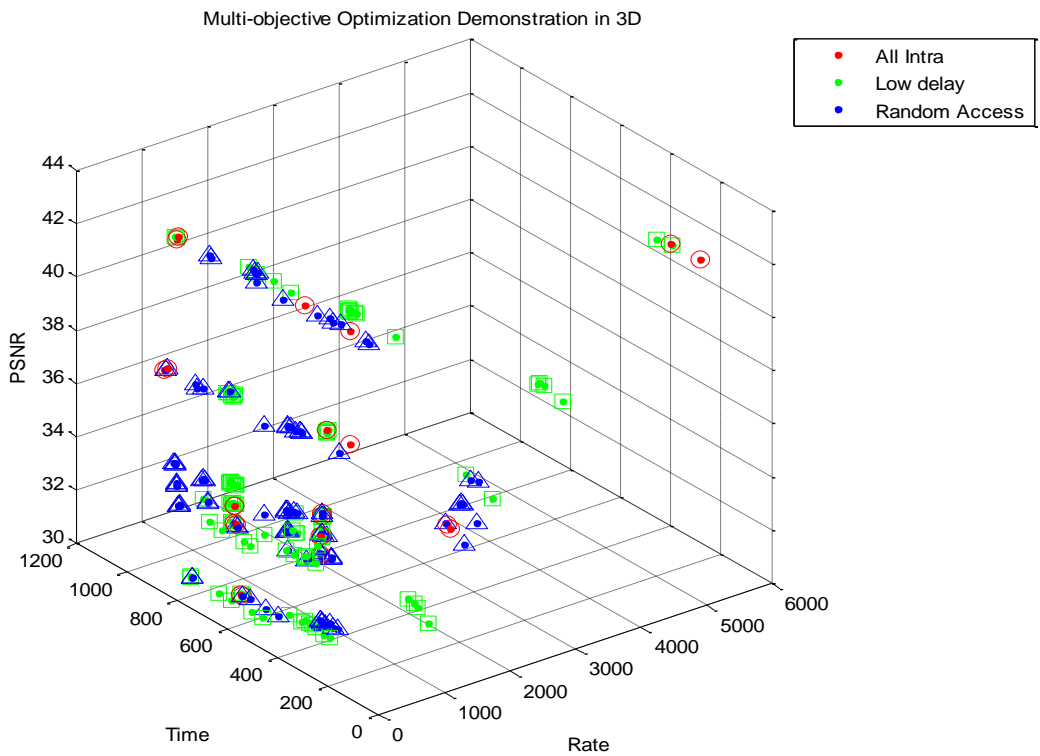

**Fig. 4.4.2** Multi-objective Optimization Space for frames 101-200 of Basketball Pass 416x240 video.

**Table 4.4.** Extended GOP configuration modes that extend the standard modes (see Table 4.1). The modes were applied to the first 100 video frames of the basketball pass video (416x240). We have a total of 216 GOP configurations. The new GOP configurations are LD6 and RA4.

| Mode | QP | SAO | DBF | Decoding Refresh | Number of Configs |
|---|---|---|---|---|---|
| All I | 22, 27, 31, 32, 33, 37 | ON/OFF | ON/OFF | - | 24 |
| RA 8 | 22, 27, 31, 32, 33, 37 | ON/OFF | ON/OFF | IDR/CDR | 48 |
| RA 4 | 22, 27, 31, 32, 33, 37 | ON/OFF | ON/OFF | IDR/CDR | 48 |
| LD 4 | 22, 27, 31, 32, 33, 37 | ON/OFF | ON/OFF | IDR/CDR | 48 |
| LD 6 | 22, 27, 31, 32, 33, 37 | ON/OFF | ON/OFF | IDR/CDR | 48 |

The new GOP configurations (LD6, RA4) are derived from the standard GOP configurations to extend the Pareto front. The goal here is to provide for a Pareto-front surface that will allow for finer DRASTIC control.

### 4.4.2.1 Pareto Front for the second 100 frames

The Pareto front for the next set of frames is constructed similar to the first encoding experiment for the minimum bitrate and minimum time modes. The Pareto front is shown in Fig. 4.4.2.1. Out of the 216 configuration points in the optimization space, we got 96 configuration points that solves for the above constraints. In addition to the standard GOPs, the new GOP structures find their place in middle of the plot with the points in the Pareto optimization space compared to fig. 4.4.1.1 where there are no points in the Pareto front.

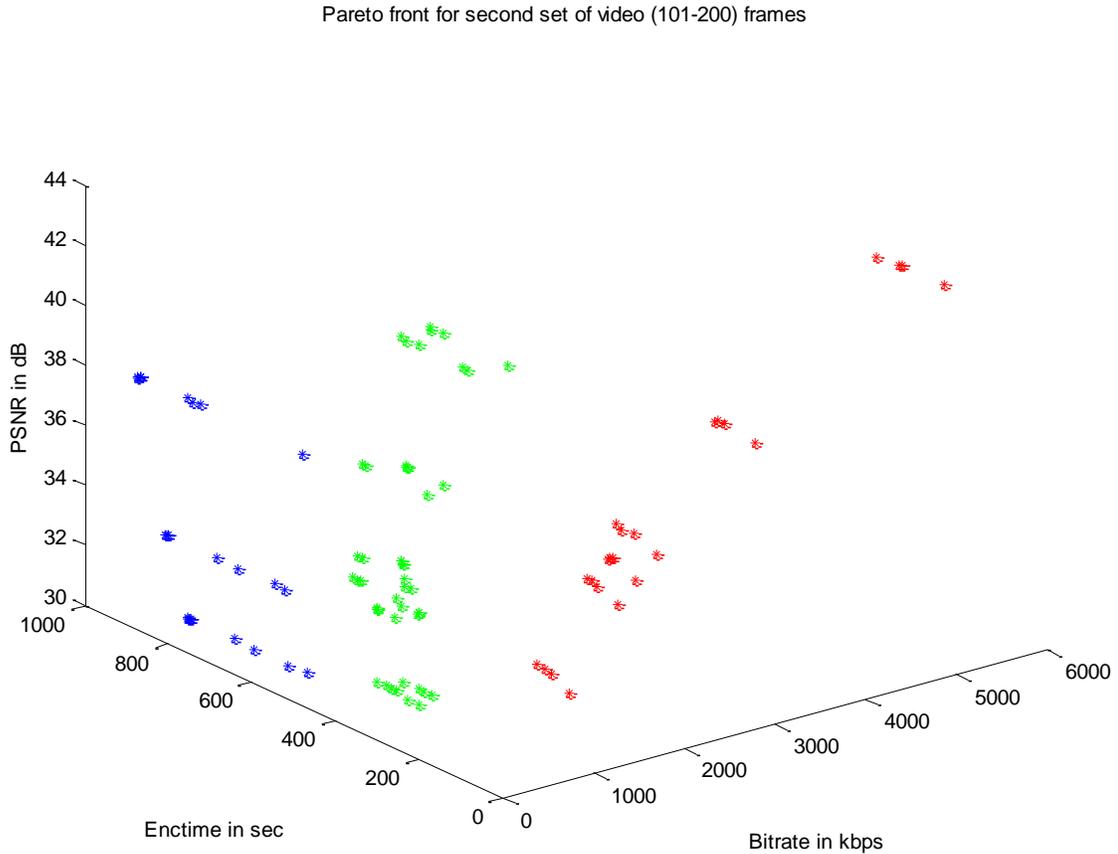

Fig.4.4.2.1 Pareto Front for the second 100 frames of Basketball Pass video (416x240)

## 4.5 Simulation Results with Switching GOP (Group of Pictures) modes

All the simulations runs were performed on a Windows 8 64-bit platform with 4GB RAM (1.6 GHz) using an AMD FX 8350 microprocessor with 8 cores (8 threads) running at 4GHz. For estimating encoding times, we use the standard reference software [reference].

## 4.5.1 Switching in Minimum Bitrate Mode

We show the Pareto front for a single HEVC video of resolution 416x240 in Fig. 4.4.2.1. We also show examples of switching among minimum bitrate modes in Fig. 4.5.1. We note the significant bitrate savings in the first 100 video frames over the second 100 video frames. These results demonstrate the advantages of using dynamic reconfiguration versus static approaches.

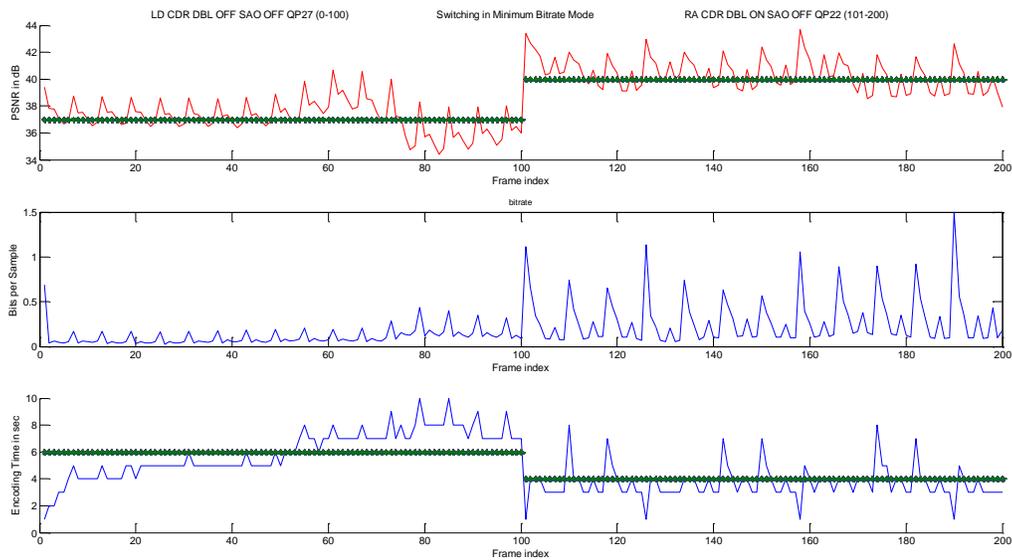

**Fig. 4.5.1.** DRASTIC mode switching for minimum bitrate mode. For the first 100 frames, we require that the encoding time remains under 600 sec and the PSNR remains above 35 dB. For the second 100 frames, the minimum PSNR level is changed to 40 dB and the maximum encoding time is changed to 360 sec.

The above plot shows the minimum bit rate savings as the video switches from the frame 100 frames over the next 100 frames. This scenario can be explained as follows. So in this switching mode, if the user wants to achieve minimum bit rate then the objective is to maintain minimum bit rate and solving for the other two constraints that is maximum encoding time and minimum acceptable level of video quality. So the encoding is done from the Pareto configurations obtained from the Pareto front that satisfy these constraints. So the user can encode the first 100 video frames with a minimum bit rate but at a longer processing time with a minimum video quality to handle the bandwidth of the device been used. And now when there is an availability of more bandwidth, the encoding of the next 100 frames switches to a better level of video quality than the first 100 frames with shorter processing time.

## 4.5.2 Switching in Maximum Video Quality Mode

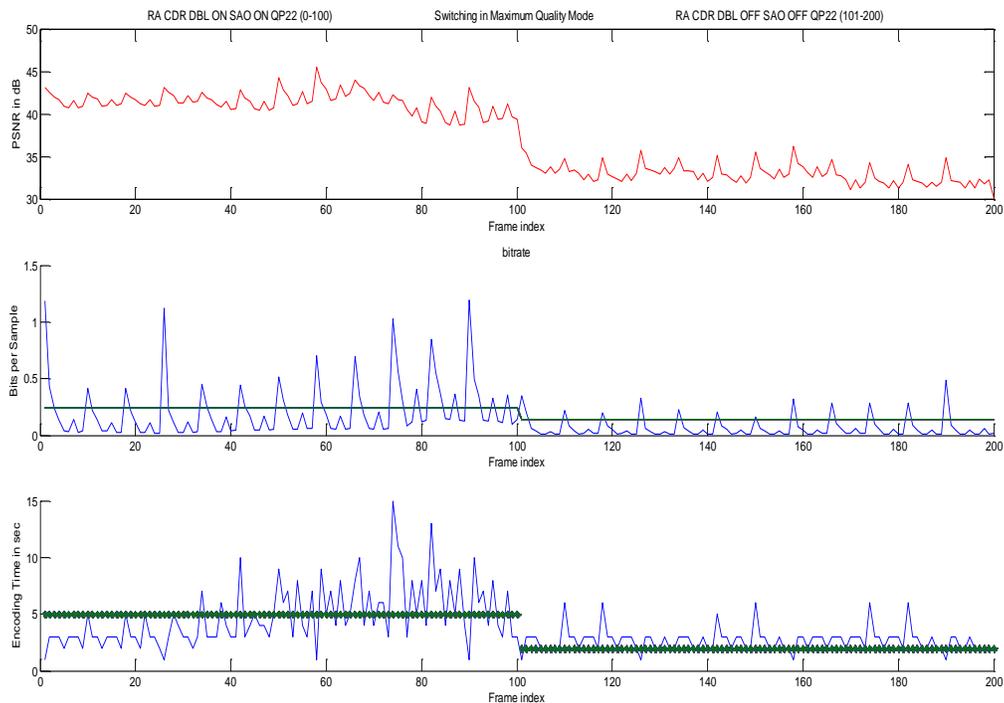

**Fig. 4.5.2.** DRASTIC mode switching in maximum quality mode. For the first 100 video frames, the requirement for the average encoding time is to be below 70 sec and the average bits per sample should remain below 500 bits per/sample. For the second 100 video frames, the requirements are for the average encoding time is 200 sec.

The switching in maximum video quality is another example which can be used in real time. Here the constraints solve for maintaining a higher video quality without exceeding the maximum bandwidth available and maximum encoding time. For an example, user encodes the first 100 frames with a high video quality where the bandwidth was high since there wasn't much data usage so the video takes a time lesser than the maximum available. Suppose when the network coverage of the device goes down, then switching will help the device to switch it to lower bit rate and still provide a better video quality.

# Chapter 5

# Relational Video Database (RVD) Formulation for Implementing DRASTIC Modes and Future Work

## 5.1 Introduction

Databases are ubiquitous and are the driving wheels behind our massive internet web of information. From powerful search engines to any small web application all the data in the web is obtained from them. From smart phones to super computers the applications of databases are diverse and are extensively used to index information from large datasets. On the other hand, in terms of storage requirements and internet traffic, the majority of the web traffic is dominated by video data.

This chapter aims to use relational databases to describe the relationships among HEVC video encoder configurations and parameters with the devices and networks that are used for communicating the videos. This is implemented in two phases where the first phase involves the design of database tables that describe encoder configurations, device screen resolutions, and network configurations. Then, a database query is used for implementing the DRASTIC modes. The advantage of the database formulation is that we can use them to retrieve optimal encoder configurations for different devices and network conditions. For example, we can query the RVD to obtain an optimal HEVC configuration for encoding a video at 1920 HDTV resolution at typical 4G network speeds. The query will return configurations that can encode videos at this specific constraint. Similarly, a high quality video can be encoded at a minimum bitrate mode using the DRASTIC table derived from the relevant tables. Essentially, instead of returning a single DRASTIC configuration, the use of relational databases allows us to retrieve Pareto fronts for different devices and network conditions.

## 5.2 Relational Video Database Model

In this relational model, the results of the Pareto optimal front and the HEVC [4] encoder configurations are mapped in terms of database tables. The RVD database tables are shown in Figs. 5.1 and 5.2 (see [20]).

*mysql> select \* from Videosource;*

| Source_Id | Resolution | framerate | uncvideoformat |
|---|---|---|---|
| SV001 | 416x240 | 30 | YUV |

*mysql> select \* from Videoseg;*

| Video_Id | Resolution | start_frame | end_frame | Source_Id |
|---|---|---|---|---|
| V001 | 416x240 | 1 | 100 | 1 |
| V002 | 416x240 | 101 | 200 | 1 |

*mysql> select \* from Softwareconfig;*

| SW_Id | QPvalue | GOPconfig | DBL | SAO |
|---|---|---|---|---|
| S1 | 22 | AI | ON | ON |
| S2 | 27 | AI | ON | ON |
| S3 | 31 | AI | ON | ON |
| ... | | | | |

*mysql> select \* from Paretofront;*

| Pareto_Id | SW_Id | Video_Id | Enc_video_id | PSNR | Enctime | Bitrate |
|---|---|---|---|---|---|---|
| P001 | S1 | V001 | EV001 | 43 | 107 | 4867 |
| P002 | S2 | V001 | EV002 | 40 | 104 | 2860 |
| P003 | S3 | V001 | EV003 | 37 | 126 | 1825 |
| ... | | | | | | |

**Fig. 5.1** Pareto Front Tables (Videosource, Videoseg, Softwareconfig, Paretofront).

The database tables are summarized below for the example considered in this thesis:

- The *videosource* table describes the source video with **Source_Id** as primary key and represents the original uncompressed video YUV format with a resolution 416x240.

- The *videoseg* table describes the segmented video 'V001' representing video frames from (1-100) and 'V002' representing frames (101-200) as start and end frames mentioned and use **video_Id** as its primary key **Basketball Pass** as source file name.

- The *Softwareconfig table* describes the different configurations. It uses **SW_Id** as its primary key. It is characterized by the QP value (QPvalue) and the GOP configuration (GOPconfig) and the filter parameters Deblocking (DBL) and Sample Adaptive Offset (SAO). The Softwareconfig table has 336 records representing the encoder configurations obtained from Encoding Experiment I [4.4.1] & Encoding Experiment II [4.4.2].

- The *Paretofront* table describes the Pareto front obtained from the various software configurations for the video segments and uses **Pareto_Id** as the primary key, **SW_Id** as the foreign key referencing the Softwareconfig table. It has additional parameters PSNR, Bitrate and Enctime obtained after running the encoder with the configurations from Softwareconfig table. It uses V001 & V002 to refer which video segment is encoded. The encoded videos are stored in filenames that are stored in the *Enc_video_id* field. The Paretofront has a total of 146 records representing the optimal Paretofront configurations obtained from [4.4.1.1] & [4.4.2.1].

*mysql> select * from Deviceconfig;*

*Dev_Id Displayresolution Maxplaybackframerate Maxencodeframerate Device_typ Networktypes*

*Nexus 5                   1920x1080    30   30   Smartphone   GSM/2G/3G/4G LTE*
*iPhone 5S Model A153   1136x640    30    30   Smartphone GSM/EDGE/LTE/HSDPA*

*mysql> select * from Network;*

| Networktype | TheorDL | TheorUL | TypDL | TypUL |
|---|---|---|---|---|
| GSM | 14.4 kbps | 14.4 kbps | 10 kbps | 10 kbps |
| 2G | 9.6 kbps | 115 kbps | 10 kbps | 10 kbps |
| 3G | 144 kbps | 2 Mbps | 220 kbps | 384 kbps |
| 4G LTE | 1 Gbps | 100 Mbps | null | null |

*mysql> select * from Userconfig;*

| UserDevProf | Dev_Id | Profile | PSNR | Enctime | Bitrate |
|---|---|---|---|---|---|
| UserDevProf 1-1-1 | Nexus 5 | low | 30 | 60.56 | 100 |
| UserDevProf 1-1-2 | Nexus 5 | medium | 35 | 125.46 | 180 |
| UserDevProf 1-1-3 | Nexus 5 | high | 40 | 232.65 | 450 |
| UserDevProf 1-2-1 | iPhone 5s Model A 15 | low | 29 | 77.4 | 95.2 |
| UserDevProf 1-2-2 | iPhone 5s Model A 15 | medium | 34.68 | 165.3 | 201.65 |
| UserDevProf 1-2-3 | iPhone 5s Model A 15 | high | 39.45 | 288.62 | 567.65 |

Fig. 5.2 Database Tables (Deviceconfig, Userconfig, Network)

- *Deviceconfig table* is used to describe the different devices that are used for encoding and decoding the video. It uses **Dev_Id** as its primary key. Each device is described by its DisplayResolution, Maxplaybackframerate, Maxencodeframerate, Device_type and Networktypes. The devices are supported in wide network coverage and each device has its own display resolution to playback the video.

- The *Network table* that is used to describe the network has **Networktype** as its primary key. Each network is described by its Theoretical Uplink and Downlink rates and the typical transfer rates used in real time communications.

- The *Userconfig* table represents the user profiles with *UserDevProf* as the primary key

as a single user can play a video in many devices with different profiles (low, medium, high).

## 5.3 DRASTIC mode implementations using database queries

We next implement DRASTIC [13] [14] modes using queries on the database tables. A straight-forward implementation of the maximum image quality mode is given by:

    *mysql> select Pareto_Id, Enc_video_id, SW_Id, MAX(PSNR), Enctime, Bitrate*
        *from Paretofront*
        *where Bitrate <= 600 AND Enctime <= 500 AND Video_Id = 'V001'*

        *Pareto_Id  Enc_video_id  SW_Id  PSNR  Enctime  Bitrate*
         *P0031       EV0031        S74    38     302      548*

    *mysql> select Pareto_Id, Enc_video_id, SW_Id, PSNR, Enctime, Bitrate*
        *from Paretofront*
        *where Bitrate <= 1000 AND Enctime <= 500 AND Video_Id = 'V002'*

        *Pareto_Id Enc_video_id SW_Id  PSNR  Enctime  Bitrate*
         *P00131    EV00131    S166   37     448     806*

Here, we note that the Pareto-optimal configuration is retrieved in SW_id. Furthermore, the average performance for this configuration will be represented in PSNR, Enctime, and Bitrate. The encoded video is also given in Enc_video_id.

The query results for encoding the video segments in the Minimum Bitrate mode are given by:

    *mysql> select Pareto_Id, Enc_video_id, SW_Id, PSNR, Enctime, MIN(Bitrate)*
        *from Paretofront*
        *where PSNR >= 40 AND Enctime <= 800 AND Video_Id = 'V001'*

        *Pareto_Id  Enc_video_id SW_Id PSNR Enctime Bitrate*
         *P001        EV001         S1    43    107     1085*

*mysql> select Pareto_Id, Enc_video_id, SW_Id, PSNR, Enctime, MIN(Bitrate)*
*from Paretofront*
*where PSNR >= 35 AND Enctime <= 800 AND Video_Id = 'V002'*

*Pareto_Id  Enc_video_id SW_Id  PSNR Enctime Bitrate*
*P0051   EV0051   S127   42   109   732*

We can also retrieve mode information based on the user profiles. For example, to select the high-profile mode for the Nexus 5, we can simply use:

*mysql> set @maxBitrate = (select Bitrate from Userconfig*
*where Dev_id='Nexus 5' and Profile='High')*

which stores a maximum bitrate of 450 Kbps. Then, set the encoding time using:

*mysql> set @maxEncTime = (select Enctime from Userconfig*
*where Dev_id='Nexus 5' and Profile='High')*

which gives a maximum encoding time of 507 seconds. We can then implement the maximum video quality mode using the retrieved values:

*mysql> select Pareto_Id, Enc_video_id, SW_Id, MAX(PSNR), Enctime, Bitrate*
*from Paretofront*
*where Bitrate <= @maxBitrate AND Enctime <= @maxEncTime*
*AND Video_Id = 'V001'*

*Pareto_Id Enc_video_id SW_Id  PSNR  Enctime  Bitrate*
*P0035    EV0035    S75    35    306    315*

These results can be modified depending on the user requirements.

## 5.4 Summary

In this chapter, a relational database formulation was used to implement the DRASTIC modes for Maximum Quality and Minimum Bitrate modes. Currently the database has encoder configurations for only a single video resolution. This idea can be further expanded and the database can have many different videos of higher resolution and the corresponding encoder configurations. Such a database will be an effective for Big Data and Internet of Things (IoT) as the future technical world will be governed by powerful and intelligent databases with so many applications from a wider perspective.

**5.5 Conclusion & Future Work**

This thesis concludes with the work we have presented by using dynamically reconfigurable encoding for meeting varying constraints imposed on the video. Ongoing research involves the development of optimized implementations to support the development of GOP parallelization from a hardware platform and to provide a minimum encoding time mode. Furthermore, statistical fitting models for the Pareto front can be developed for real-time estimation and automatic constraint generation based on the incoming video.